\documentclass[12pt,a4paper]{article}

\usepackage[T1]{fontenc}
\usepackage[utf8]{inputenc}
\usepackage{lmodern}
\usepackage[margin=2.5cm]{geometry}
\usepackage{setspace}
\usepackage{amssymb}
\usepackage{amsmath}

\usepackage{siunitx}
\usepackage{graphicx}
\usepackage{booktabs}
\usepackage{float}
\usepackage[font=small,labelfont=bf]{caption}

\usepackage{microtype}
\usepackage{ragged2e}
\usepackage{textcomp}

\usepackage{xcolor}
\definecolor{linkblue}{HTML}{0000CC}
\definecolor{orcidgreen}{HTML}{A6CE39}

\usepackage[colorlinks=true,
linkcolor=linkblue,
citecolor=linkblue,
urlcolor=linkblue,
filecolor=linkblue,
pdfstartview=FitH,
breaklinks=true,
pdftitle={Electron-Impact Quasi-Resonant Ion-Pair Dissociation of OCS},
pdfauthor={Narayan Kundu, Soumya Ghosh, Dhananjay Nandi},
pdfsubject={Chemical Physics, Molecular Physics}
]{hyperref}

\usepackage[numbers,compress,sort&compress]{natbib}
\newcommand{\OCS}{\text{OCS}}
\newcommand{\COp}{\text{CO}^{+}}
\newcommand{\CSp}{\text{CS}^{+}}
\newcommand{\OSp}{\text{OS}^{+}}
\newcommand{\COm}{\text{CO}^{-}}
\newcommand{\CSm}{\text{CS}^{-}}
\newcommand{\OSm}{\text{OS}^{-}}
\newcommand{\Sm}{\text{S}^{-}}
\newcommand{\Om}{\text{O}^{-}}
\newcommand{\Cm}{\text{C}^{-}}
\newcommand{\Sp}{\text{S}^{+}}
\newcommand{\Op}{\text{O}^{+}}
\newcommand{\Cp}{\text{C}^{+}}
\newcommand{\OCSm}{\text{OCS}^{-}}
\newcommand{\OCSss}{\text{OCS}^{**}}
\newcommand{\Otwo}{\text{O}_{2}}
\newcommand{\CO}{\text{CO}}
\newcommand{\CS}{\text{CS}}
\newcommand{\Ap}{\text{A}^{+}}
\newcommand{\Bm}{\text{B}^{-}}



\title{%
	\textbf{%
		Electron-Impact Quasi-Resonant Ion-Pair Dissociation of OCS:
		A Velocity Slice Imaging Study with Partial Wave Analysis%
	}%
}

\author{%
	Narayan Kundu$^{1,2}$\thanks{E-mail: \texttt{kundu.narayan1995@gmail.com}},\quad
	Soumya Ghosh$^{1}$,\quad
	Dhananjay Nandi$^{1,3}$\thanks{E-mail: \texttt{dhananjay@iiserkol.ac.in}}
	\\[2ex]
	\small
	\begin{minipage}{\textwidth}
		\centering
		$^{1}$Department of Physical Sciences,
		Indian Institute of Science Education and Research Kolkata,\\
		Mohanpur--741246, India
		\\[0.5ex]
		$^{2}$University of Kassel, Institute of Physics,
		Heinrich-Plett-Str.\ 40, 34132 Kassel, Germany
		\\[0.5ex]
		$^{3}$Center for Atomic, Molecular and Optical Sciences
		and Technologies,\\
		Joint initiative of IIT Tirupati and IISER Tirupati,
		Yerpedu--517619, India
	\end{minipage}%
}

\date{\today}

\begin{document}
	
	\maketitle
	
	\begin{abstract}
		\noindent
		We present velocity map imaging data on intramolecular ion-pair dissociation (IPD) of carbonyl sulfide ($\OCS$) induced by electron impact over the \SIrange{20}{45}{\eV} energy range. Two distinct IPD pathways were resolved: $\COp + \Sm$ (threshold $14.8 \pm 0.7$~eV) and $\CSp + \Om$ (threshold $16.8 \pm 0.7$~eV). The kinetic energy release spectra display a single peak for $\Sm$ but split into two components for $\Om$; in both channels the maximum kinetic energies level off once the beam energy exceeds roughly \SI{30}{\eV}, pointing to excitation through discrete superexcited states of quasi-resonant character. Partial wave decomposition of the fragment angular distributions reveals that the momentum-transfer parameter $\beta$ surpasses unity at every energy studied, invalidating the dipole-Born approximation, and that the dominant partial wave character shifts systematically with beam energy. These patterns are consistent with a mechanism in which the incident electron deposits energy through inelastic scattering, populating hybrid Rydberg--ion-pair superexcited configurations that subsequently undergo state-specific unimolecular dissociation along nonadiabatic pathways. From an applied standpoint, intramolecular ion-pair dissociation matters for astrochemistry and radiation biophysics because it generates reactive anions and cations without photon emission, redistributing excess molecular energy nonadiabatically in environments ranging from interstellar clouds to biological systems.
	\end{abstract}
	
	
	\bigskip
	\sloppy
	
	\section{Introduction}
	
	Ion-pair dissociation (IPD)---the fragmentation of a neutral molecule into a cation--anion pair---provides a window onto long-range electron correlation, intramolecular charge migration, and the coupling between bound electronic configurations and the dissociative continuum \cite{suits2006ion,kirrandar_pra,kirrander2018heavy,kirrander_ipd_ungrade}. When fragmentation proceeds through discrete intermediate levels rather than by direct excitation into the continuum, the product-state distributions carry an imprint of those resonant configurations, and the process acquires a \textit{quasi-resonant} character \cite{bardsley1968resonant,o1968angular}. We use ``quasi-resonant'' to mean that IPD is mediated by discrete doorway states of finite width, in contrast to direct dissociation into the ion-pair continuum.
	
	Unlike ordinary dissociation routes that release neutral or singly charged fragments, IPD passes through electronically excited configurations where the two incipient ionic fragments remain transiently linked by their mutual Coulomb attraction \cite{dressed_ipd_prl,greene2000creation}. Such transient configurations emerge at avoided crossings between diabatic ionic and covalent curves; strong electronic coupling at these crossings reshuffles the bare diabatic states into adiabatic eigenstates of mixed character \cite{kirrander2018heavy,rost_prl_dressed,ulrms_molphy,kirrandar_pra}.
	
	These mixed states connect to the concept of \textit{heavy Rydberg} systems---deeply bound vibrational levels held together by the long-range Coulomb well of an ion pair \cite{kirrander2018heavy,fermi1934sopra,kirrandar_pra,hrs_review}. For a diatomic pair $\Ap\,\Bm$ the effective internuclear potential reads (in atomic units)
	\begin{equation}
		V^{\text{ion}}(R) = D_{\text{A}^{+}\text{B}^{-}} - \frac{1}{R} - \frac{\alpha_{\text{A}^{+}} + \alpha_{\text{B}^{-}}}{2R^{4}},
		\label{eq:ion_pair_potential}
	\end{equation}
	with $R$ the internuclear separation, $D_{\text{A}^{+}\text{B}^{-}}$ the ion-pair dissociation energy, and $\alpha_{\text{A}^{+}}$, $\alpha_{\text{B}^{-}}$ the static dipole polarisabilities \cite{kirrander2018heavy,hrs_review}. The leading $-1/R$ Coulomb tail supports vibrational levels whose energies obey a Rydberg-type formula:
	\begin{equation}
		E_n = D_{\text{A}^{+}\text{B}^{-}} - \frac{R_{\infty} h c \,\mu_{\text{AB}}/m_e}{(n - \delta)^{2}},
		\label{eq:heavy_rydberg_formula}
	\end{equation}
	where $n$ labels the vibrational quantum number, $\delta$ is a quantum defect, $\mu_{\text{AB}}$ is the reduced mass, $m_e$ the electron mass, $R_{\infty}$ the Rydberg constant, $h$ Planck's constant, and $c$ the speed of light \cite{kirrander2018heavy,hrs_review}. Although these expressions are written for a diatomic ion pair, they serve as a conceptual framework for the triatomic case, where the internal degrees of freedom of the diatomic fragment introduce additional structure.
	
	Populating ion-pair states by electron impact requires first lifting the neutral molecule into a \textit{superexcited state}---an electronic configuration whose energy sits above the lowest ionisation threshold yet which retains discrete character because it couples to closed dissociation channels \cite{hatano2001interaction,platzman1962superexcited,hao2017superexcited,Kouchi2013superexcited,jungen2019molecular}. When a superexcited state carries appreciable Rydberg character, its outer electron orbits at large radius while the ionic core begins to fragment \cite{merkt1997molecules,merkt1998high}. Configuration mixing between the bare ionic state $|\text{A}^{+}\text{B}^{-}\rangle$ and Rydberg configurations yields \textit{hybrid Rydberg--ion-pair states}:
	\begin{equation}
		|\Psi_{\text{hybrid}}\rangle = c_0\,|\text{A}^{+}\text{B}^{-}\rangle + \sum_{n\ell} c_{n\ell}\, |\text{A-B}^{*};\,n\ell\rangle,
		\label{eq:hybrid_state}
	\end{equation}
	with mixing coefficients set by the energy offset from the relevant Rydberg series and by the electronic coupling matrix elements \cite{martin1997electric,wehrli2021charge_pre}. The resulting adiabatic states have shifted energies, altered dissociation limits, and angular distributions reflecting the partial-wave composition of the Rydberg component \cite{rost_prl_dressed,greene2000creation}.
	
	Once a superexcited state is reached, its fate hinges on \textit{nonadiabatic coupling} between ionic and covalent surfaces. As the fragments separate along the repulsive ion-pair curve, the system can hop between diabatic surfaces at avoided crossings via Landau--Zener transitions \cite{zener1932non,worth2004beyond,yarkony2012nonadiabatic}:
	\begin{equation}
		P_{\text{LZ}} = \exp\!\left( -\frac{2\pi\,|V_{12}|^{2}}{\hbar\,v\,|\Delta F|} \right),
		\label{eq:landau_zener}
	\end{equation}
	where $V_{12}$ is the diabatic coupling element, $v$ the nuclear speed at the crossing, and $\Delta F$ the slope difference of the two diabatic curves \cite{zener1932non,worth2004beyond}. Measuring the kinetic energy release (KER) gives direct experimental access to these branching dynamics \cite{suits2006ion,berkowitz2012photoabsorption}.
	
	\begin{figure}[tb]
		\centering
		\includegraphics[width=0.85\textwidth]{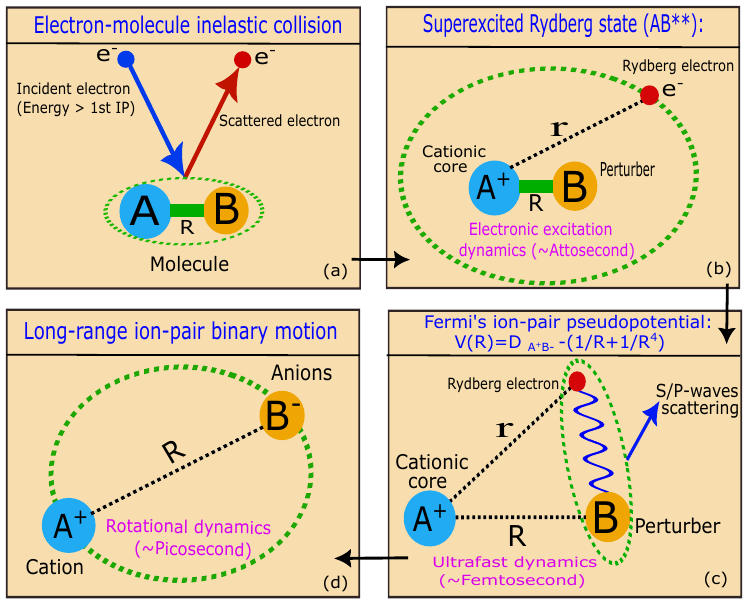}
		\caption{%
			Schematic illustration of quasi-resonant intramolecular ion-pair formation via superexcited Rydberg states following electron--molecule inelastic collisions. (a)~An incident electron with energy exceeding the first ionisation potential (IP) collides inelastically with a molecule AB, producing a scattered electron and depositing energy into the target. (b)~The collision prepares a superexcited Rydberg state ($\text{AB}^{**}$) consisting of a cationic core and a weakly bound Rydberg electron; coupling between the Rydberg manifold and the ion-pair continuum yields hybrid states. Electron impact can also access optically dark states through exchange scattering. (c)~Within the molecular framework, the Rydberg electron interacts with the evolving ion-pair core---this panel draws on concepts from heavy Rydberg physics \cite{kirrander2018heavy,fermi1934sopra} as a conceptual analogy rather than a literal description at equilibrium geometry. (d)~The ion pair $\Ap + \Bm$ separates along the intramolecular dissociation coordinate with Coulomb-driven binary motion.%
		}
		\label{fig:dressed_ipd_scheme}
	\end{figure}
	
	Electron-impact excitation offers a practical advantage over photoexcitation: it can reach \textit{optically dark states} that lack a dipole transition moment from the ground state but become accessible through exchange scattering or higher multipole interactions \cite{larsen_dark,khakoo2005dark,compton1968_dark,simpson1965_dark_ebeam}. For $\OCS$, the $a\,^{3}\Pi$ and $A'\,^{3}\Sigma^{+}$ triplet states may funnel population toward ion-pair continua through intersystem crossing \cite{suzuki1998nonadiabatic,marian2012spin,greene2000creation,rothbaum2026ultrafast}.
	
	Carbonyl sulfide ($\OCS$) is an attractive target for probing intramolecular IPD. As a linear triatomic with $C_{\infty v}$ symmetry, it supports competing exit channels:
	\begin{align}
		\OCS + e^{-} &\to \COp + \Sm + e^{-}, \label{eq:ipd_s}\\
		\OCS + e^{-} &\to \CSp + \Om + e^{-}. \label{eq:ipd_o}
	\end{align}
	Photoelectron work has mapped out the electronic structure of $\OCS$ \cite{mcglynn1971electronic}, revealing multiple Rydberg series converging to ionic thresholds \cite{suzuki1998nonadiabatic}. Mitsuke and co-workers determined ion-pair onset energies using photoionisation \cite{mitsuke1993negative}, while earlier electron-impact work charted dissociative electron attachment (DEA) resonances below 10~eV \cite{iga1995dissociative,ziesel1975s,abouaf1976}. In a recent study we measured absolute cross sections for $\Sm$ yield across both regimes~\cite{Ghosh_ocs_2025} and found that bent nuclear geometries contribute significantly to the dynamics. What has been missing is a kinematically complete measurement---velocity-resolved product distributions---for electron-impact IPD of $\OCS$. Such data also matter because $\OCS$ is the most abundant sulfur-bearing trace gas in the stratosphere \cite{ueno_young_sun_para,hattori2020constraining}.
	
	Here we use velocity map imaging (VMI) paired with time-of-flight mass spectrometry to map out the kinematics of electron-impact quasi-resonant IPD of $\OCS$ for beam energies between 20 and 45~eV \cite{eppink1997velocity,whitaker2003imaging}. By carrying out a partial-wave decomposition that goes beyond the dipole-limited Van~Brunt formalism \cite{van1974breakdown,van1970angular}, we reproduce the measured angular correlations quantitatively. The threshold energetics, the plateau in maximum KER, and the evolving partial-wave composition build a coherent case for a quasi-resonant pathway in which hybrid Rydberg--ion-pair superexcited states are populated and then fall apart nonadiabatically.
	
	\section{Experimental methods}
	
	Our apparatus consists of a pulsed electron gun, an effusive molecular-beam inlet, a three-field velocity map imaging electrode stack, and a two-dimensional position-sensitive detector, all enclosed in a differentially pumped ultrahigh-vacuum chamber maintained at a base pressure of $5 \times 10^{-7}$~mbar \cite{nag2015complete,nandi2005velocity}.
	
	Electrons are produced by thermionic emission from a resistively heated tungsten filament and delivered in 100~ns pulses at repetition rates up to 10~kHz. Beam energy is calibrated against the 6.5~eV resonance in dissociative electron attachment to $\Otwo$ \cite{nandi2005velocity}; the energy spread is approximately 0.5--0.7~eV full width at half maximum.
	
	$\OCS$ gas (purity $>$99.9\%) enters the chamber as a room-temperature effusive beam through a 1~mm aperture. Under these conditions dimer formation is negligible. The molecular beam crosses the electron beam at $90^{\circ}$ inside the extraction region.
	
	Negative-ion products are extracted by a 3~$\mu$s pulse firing 100~ns after the electron pulse clears the interaction zone. We verify velocity-mapping by imaging $\Om$ from DEA of $\Otwo$ at known kinetic energies \cite{nag2015complete,nandi2005velocity}. Detection relies on three Z-stacked microchannel plates backed by a RoentDek HEX delay-line hexanode \cite{jagutzki2002multiple}.
	
	Recovering three-dimensional velocity information from two-dimensional images presents challenges \cite{whitaker2003imaging}. Parallel slicing \cite{gebhardt2001slice} overweights ions with small perpendicular velocity components \cite{moradmand_wedge,kundu2021effect}. Moreover, inverse Abel inversion assumes cylindrical symmetry, which fails for electron-impact IPD where $\beta$ can exceed unity and partial-wave interference introduces forward-backward asymmetries \cite{kundu2023breakdown,kundu2021effect,tomographic_vmi,tpx3_vmi}. We use a conical time-gated wedge slicing scheme that suppresses pathological overcounting while retaining angular and energetic detail \cite{kundu2023breakdown,moradmand_wedge,kundu2021effect,lin2003application}.
	
	\section{Results and discussion}
	
	\begin{figure}[htb]
		\centering
		\includegraphics[width=0.95\textwidth]{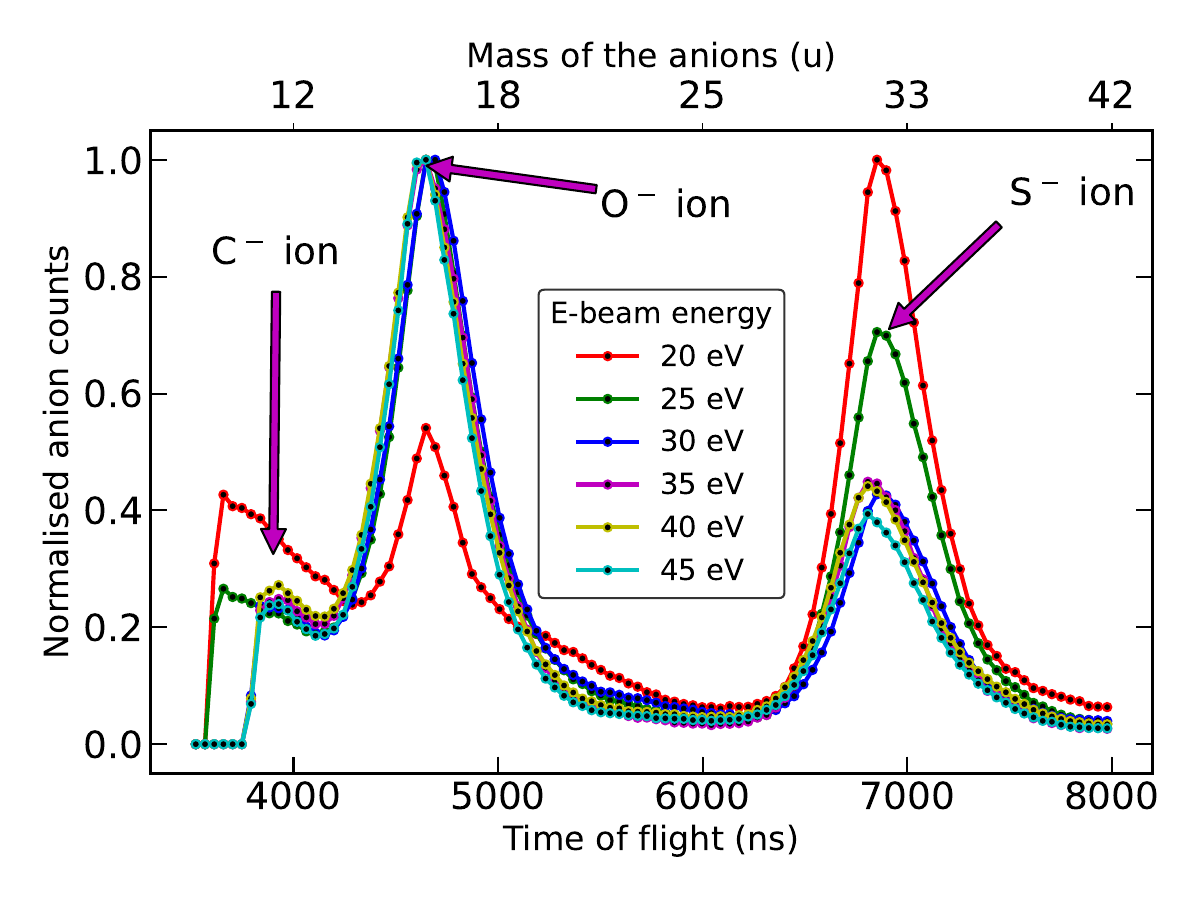}
		\caption{%
			Time-of-flight mass spectra of anionic products from electron-impact ion-pair dissociation of $\OCS$ at the incident energies indicated. The upper abscissa shows the corresponding mass-to-charge ratio. The dominant peaks correspond to $\Om$ (16~u) and $\Sm$ (32~u).%
		}
		\label{fig:tof_mass_ocs}
	\end{figure}
	
	\subsection{Time-of-flight mass spectra and identification of IPD channels}
	
	Figure~\ref{fig:tof_mass_ocs} displays time-of-flight mass spectra of negative ions generated by electron-impact IPD of $\OCS$ at beam energies spanning 20--45~eV. Two anionic species are unambiguously identified: $\Om$ ($m/z = 16$) and $\Sm$ ($m/z = 32$) \cite{wiley1955time}. A handful of counts appear near $m/z = 12$ ($\Cm$) but remain at or below the noise floor.
	
	Six ion-pair exit channels are in principle open:
	\begin{equation}
		\begin{aligned}
			\OCS + e^{-} &\rightarrow \OCSss + e^{-\prime} \\
			&\rightarrow
			\begin{cases}
				\COp + \Sm  + e^{-\prime} & \text{(Channel~I)}   \\
				\CSp + \Om  + e^{-\prime} & \text{(Channel~II)}  \\
				\OSp + \Cm  + e^{-\prime} & \text{(Channel~III)} \\
				\CSm + \Op  + e^{-\prime} & \text{(Channel~IV)}  \\
				\COm + \Sp  + e^{-\prime} & \text{(Channel~V)}   \\
				\OSm + \Cp  + e^{-\prime} & \text{(Channel~VI)}
			\end{cases}
		\end{aligned}
		\label{eq:channels}
	\end{equation}
	where $e^{-\prime}$ denotes the scattered electron. Only Channels~I and~II yield measurable signal. Channel~III is suppressed by the modest electron affinity of carbon and the need to rupture both bonds simultaneously. Channels~IV--VI require molecular anion formation in antibonding configurations and evidently cannot compete.
	
	At 20~eV the $\Sm$ count exceeds $\Om$; this ranking reverses above 25~eV. This energy-dependent branching ratio tracks the different cross-section curves of the electronic excitations feeding each channel.
	
	\subsection{Excitation functions and threshold energetics}
	
	\begin{figure}[htb]
		\centering
		\includegraphics[width=0.95\textwidth]{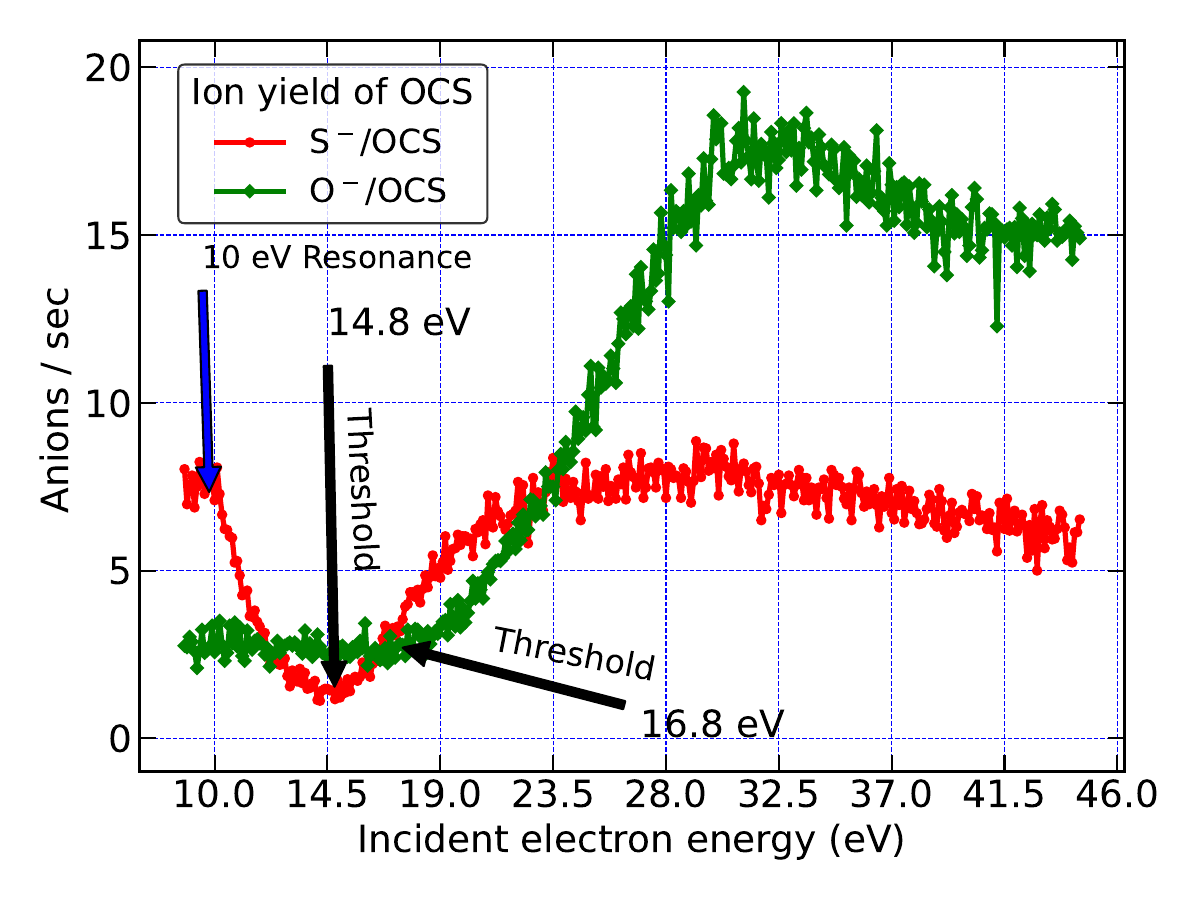}
		\caption{%
			Expanded view of the excitation functions for $\Om$ and $\Sm$ formation in the ion-pair dissociation (IPD) threshold region. Black arrows indicate the experimentally determined threshold energies for two IPD channels, and a blue arrow is used to reflect the DEA resonance of OCS near 10~eV electron beam energies \cite{kundu2024observation}.%
		}
		\label{fig:yield_fit_ocs}
	\end{figure}
	
	Figures~\ref{fig:yield_fit_ocs} plot the excitation functions for $\Om$ and $\Sm$ across 10--45~eV. Near 10~eV, structured resonance features mark dissociative electron attachment through transient $\OCSm$ \cite{kundu2024observation,iga1995dissociative,ziesel1975s,abouaf1976}. Above 14~eV, yields climb steadily, signalling IPD through superexcited states \cite{Ghosh_ocs_2025,platzman1962superexcited,kundu2023breakdown}.
	
	We extract appearance energies of $E_{\mathrm{th}}^{\mathrm{exp}}(\Sm) = 14.8 \pm 0.7$~eV for Channel~I and $E_{\mathrm{th}}^{\mathrm{exp}}(\Om) = 16.8 \pm 0.7$~eV for Channel~II (figure~\ref{fig:yield_fit_ocs}). These values are consistent with our previous report~\cite{Ghosh_ocs_2025}, studied using a different setup.
	
	The thermochemical onset for Channel~I follows from
	\begin{equation}
		E_{\mathrm{th}}^{\mathrm{calc}}(\COp + \Sm) = D_0(\mathrm{CO{-}S}) + \mathrm{IP}(\CO) - \mathrm{EA}(\text{S}),
		\label{eq:threshold_S}
	\end{equation}
	with $D_0(\mathrm{CO{-}S}) = 3.21$~eV \cite{darwent1970dissociation}, $\mathrm{IP}(\CO) = 14.014$~eV \cite{erman1993direct}, and $\mathrm{EA}(\text{S}) = 2.077$~eV \cite{blondel2001electron,ea_sulfur}, giving $E_{\mathrm{th}}^{\mathrm{calc}}(\Sm) = 15.15$~eV. This agrees with our measured value and with Mitsuke \textit{et al.}~\cite{mitsuke1993negative}.
	
	For Channel~II:
	\begin{equation}
		E_{\mathrm{th}}^{\mathrm{calc}}(\CSp + \Om) = D_0(\mathrm{O{-}CS}) + \mathrm{IP}(\CS) - \mathrm{EA}(\text{O}),
		\label{eq:threshold_O}
	\end{equation}
	with $D_0(\mathrm{O{-}CS}) = 6.52$~eV \cite{darwent1970dissociation}, $\mathrm{IP}(\CS) = 11.33$~eV \cite{erman1993direct}, and $\mathrm{EA}(\text{O}) = 1.461$~eV \cite{blondel2001electron}, yielding $E_{\mathrm{th}}^{\mathrm{calc}}(\Om) = 16.39$~eV, consistent with our measurement and with Mitsuke \textit{et al.}~\cite{mitsuke1993negative}.
	
	\subsection{Velocity slice images}
	
	\begin{figure}[htb]
		\centering
		\includegraphics[width=0.99\textwidth]{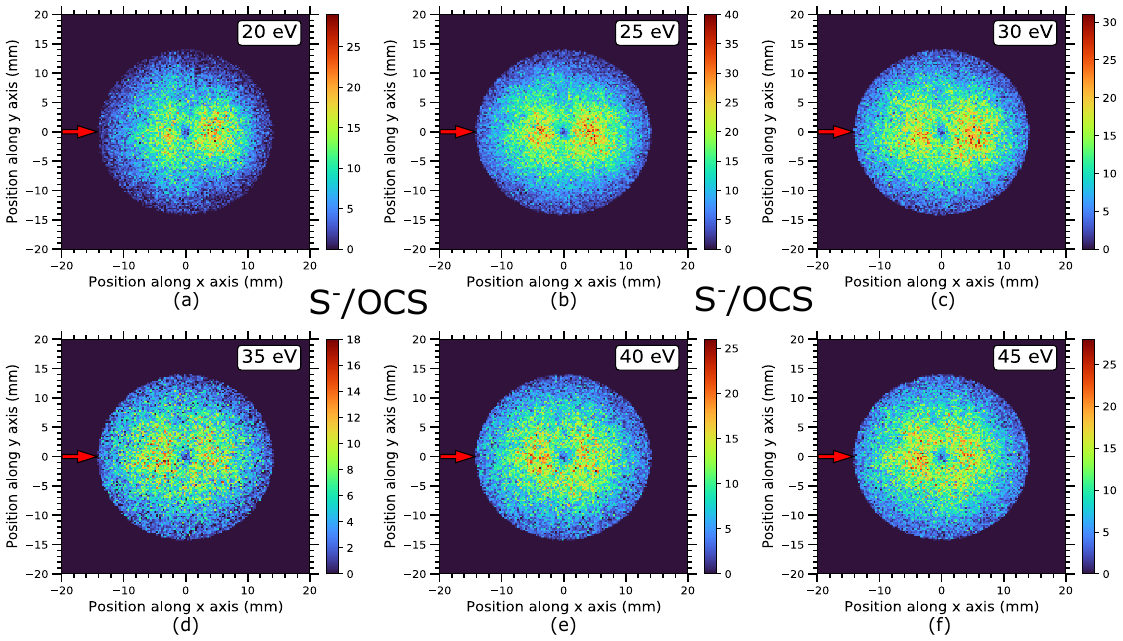}
		\caption{%
			Conical wedge slice images (50~ns gate) of $\Sm$ fragments from electron-impact IPD of $\OCS$ at (a)~20, (b)~25, (c)~30, (d)~35, (e)~40, and (f)~45~eV. Red arrows indicate the electron beam direction.%
		}
		\label{fig:allvsi_s_ocs}
	\end{figure}
	
	\begin{figure}[htb]
		\centering
		\includegraphics[width=0.99\textwidth]{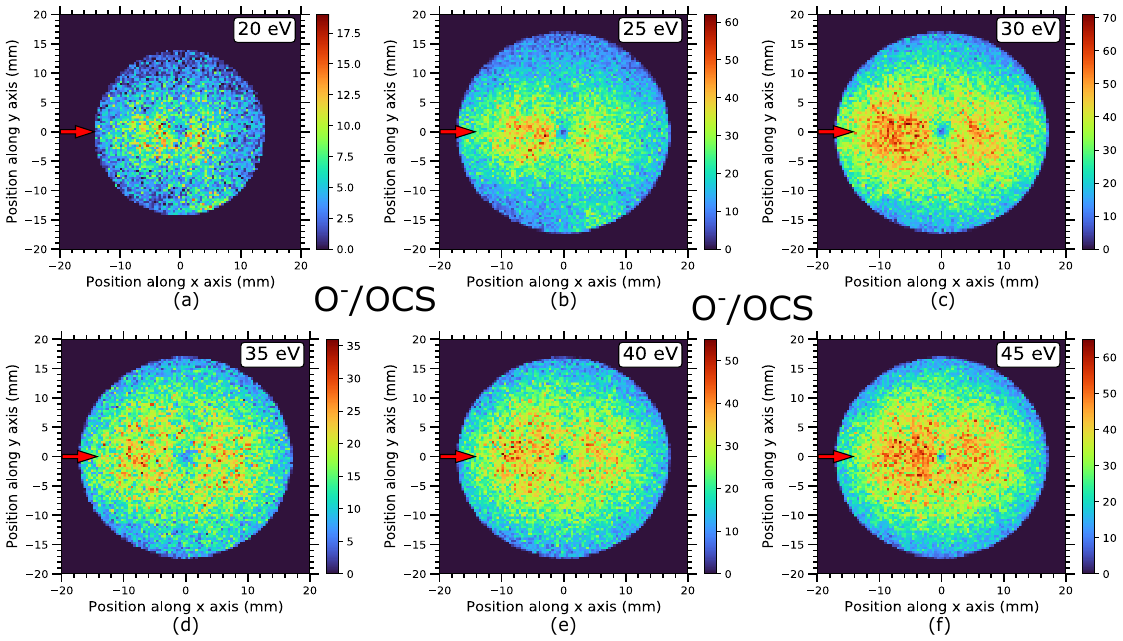}
		\caption{%
			Conical wedge slice images (50~ns gate) of $\Om$ fragments from electron-impact IPD of $\OCS$ at (a)~20, (b)~25, (c)~30, (d)~35, (e)~40, and (f)~45~eV. Red arrows indicate the electron beam direction.%
		}
		\label{fig:allvsi_o_ocs}
	\end{figure}
	
	Figures~\ref{fig:allvsi_s_ocs} and~\ref{fig:allvsi_o_ocs} show conical wedge slice images for $\Sm$ and $\Om$ at beam energies of 20--45~eV. Radial distance scales with fragment momentum, while polar angle encodes ejection direction from the beam axis \cite{eppink1997velocity,whitaker2003imaging}.
	
	Neither set of images contains sharp concentric rings that would indicate population of individual rovibronic product levels \cite{zare1967dissociation,jonah1971photolysis}. The smooth radial falloff is characteristic of predissociation through strongly coupled avoided crossings \cite{schinke1993photodissociation,kundu2022_ipd_co,kundu2023breakdown}. $\Om$ images extend to larger radii than $\Sm$ images, reflecting higher maximum kinetic energy release in Channel~II.
	
	\subsection{Kinetic energy release distributions}
	
	\begin{figure}[htb]
		\centering
		\includegraphics[width=0.95\textwidth]{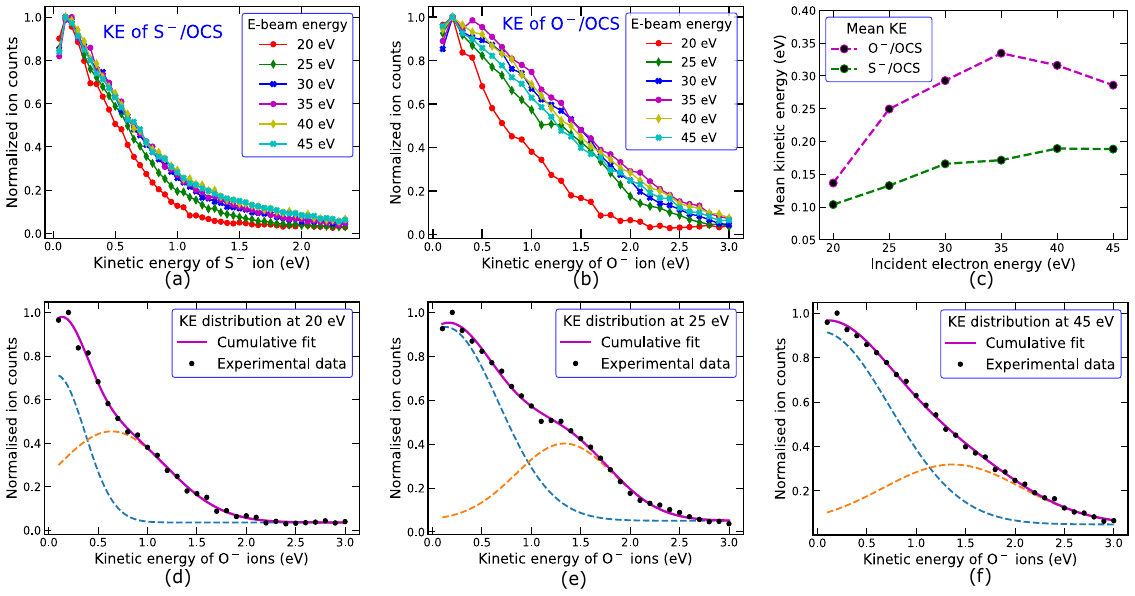}
		\caption{%
			(a,b)~Normalised kinetic energy distributions for $\Sm$ and $\Om$ fragments. (c)~Mean kinetic energy as a function of incident energy. (d--f)~Two-Gaussian decomposition of $\Om$ distributions demonstrating bimodal character.%
		}
		\label{fig:ke_ocs}
	\end{figure}
	
	Figure~\ref{fig:ke_ocs} shows kinetic energy (KE) spectra derived via Jacobian-weighted radial integration \cite{kundu2023breakdown,whitaker2003imaging}.
	
	Channel~I ($\Sm + \COp$) produces a single broad peak at every beam energy (figure~\ref{fig:ke_ocs}a), with maximum KE approximately 2.0~eV. Channel~II ($\Om + \CSp$) gives a two-humped profile (figure~\ref{fig:ke_ocs}b). Two-Gaussian fitting (figures~\ref{fig:ke_ocs}d--f) separates a slow component ($\sim$0.5--1.0~eV) from a fast one ($\sim$1.5--2.5~eV), the latter reaching approximately 3.0~eV. Two humps imply at least two distinguishable fragmentation routes \cite{levine2009molecular}.
	
	That $\Om$ ceiling KE exceeds $\Sm$ ceiling KE follows from momentum conservation: the atomic anion carries a fraction $m_{\mathrm{cation}}/(m_{\mathrm{cation}} + m_{\mathrm{anion}})$, evaluating to 0.73 for $\Om$ but only 0.47 for $\Sm$.
	
	Figure~\ref{fig:ke_ocs}c tracks mean KE against beam energy. Although the average rises steadily, the \textit{maximum} KE plateaus once the beam exceeds approximately 30~eV. This plateau is the signature of quasi-resonant IPD: fragmentation runs through discrete superexcited doorway states whose fixed energies cap the available nuclear kinetic energy \cite{merkt1997molecules,kirrander2018heavy,worth2004beyond,yarkony2012nonadiabatic,bardsley1968resonant}.
	
	\subsection{Interpretation within the heavy Rydberg framework}
	
	The two-component KE spectrum of $\Om$ has a natural interpretation within the heavy Rydberg framework. The roughly 1.2--1.5~eV gap in total KER between the two humps matches the separation expected if $\CSp$ is formed in two different electronic states:
	\begin{itemize}
		\item Slow component: $\CSp$($X\,^{2}\Sigma^{+}$) $+$ $\Om$
		\item Fast component: $\CSp$($A\,^{2}\Pi$) $+$ $\Om$
	\end{itemize}
	The $A$--$X$ interval of $\CSp$ ($\sim$2.0~eV) \cite{huber1979constants,coxon1992cs} is consistent with this assignment.
	
	In the heavy-Rydberg picture, the hybrid superexcited states of $\OCS$ carry ion-pair character correlating with different $\CSp$ electronic levels. Rydberg--ion-pair mixing (equation~\ref{eq:hybrid_state}) proceeds independently for configurations leading to each asymptote, so two separate doorway manifolds exist \cite{kirrander2018heavy,greene2000creation}.
	
	The single-peaked distribution of $\Sm$ indicates that $\COp$ emerges predominantly in its ground $X\,^{2}\Sigma^{+}$ state; the $A\,^{2}\Pi$ level ($T_e = 2.53$~eV) \cite{huber1979constants} is either too high or too weakly coupled.
	
	\subsection{Angular distributions and partial wave analysis}
	
	Fragment yield at angle $\theta$ from the beam axis follows the partial-wave expansion of Van Brunt and Kieffer \cite{van1974breakdown,o1968angular}:
	\begin{equation}
		I(\theta) = \sum_{|\mu|} \left| \sum_{l=|\mu|}^{l_{\mathrm{max}}} a_l\, i^l \sqrt{\frac{(2l+1)(l-|\mu|)!}{(l+|\mu|)!}}\, j_l(\beta)\, P_l^{|\mu|}(\cos\theta)\, e^{i\delta_l} \right|^{2}
		\label{eq:vanbrunt}
	\end{equation}
	Here $\mu = \Lambda_f - \Lambda_i$ fixes the lowest allowed partial wave, $a_l$ are partial-wave amplitudes, $j_l(\beta)$ are spherical Bessel functions evaluated at $\beta = K\bar{\gamma}$ ($K$: momentum transfer; $\bar{\gamma}$: effective molecular size), $P_l^{|\mu|}(\cos\theta)$ are associated Legendre polynomials, and $\delta_l$ are phase shifts \cite{brunger2002electron}.
	
	The parameter $\beta$ serves as a diagnostic: $\beta \ll 1$ recovers the dipole Born limit, whereas $\beta \gtrsim 1$ means several partial waves contribute \cite{kundu2023breakdown,brunger2002electron,bardsley1968resonant}.
	
	\begin{figure}[htb]
		\centering
		\includegraphics[width=0.95\textwidth]{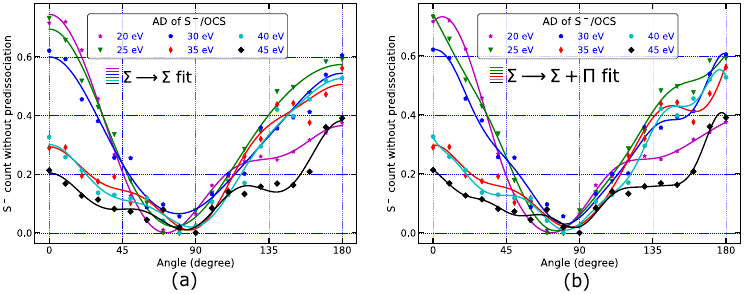}
		\caption{%
			Kinetic-energy-integrated angular distributions for $\Sm$ fragments. Solid curves are fits to equation~(\ref{eq:vanbrunt}) assuming (a)~pure $\Sigma \rightarrow \Sigma$ ($\mu = 0$) transitions and (b)~mixed $\Sigma \rightarrow \Sigma + \Pi$ transitions.%
		}
		\label{fig:ad_s_ocs_all}
	\end{figure}
	
	\begin{figure}[htb]
		\centering
		\includegraphics[width=0.95\textwidth]{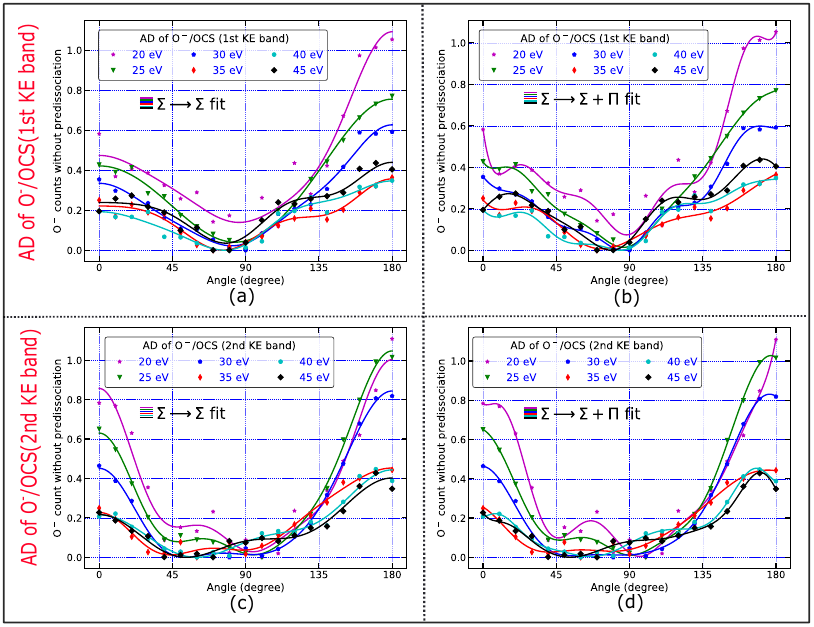}
		\caption{%
			Kinetic-energy-integrated angular distributions for $\Om$ fragments. Panels (a,c) show fits for pure $\Sigma \rightarrow \Sigma$ transitions; panels (b,d) include $\Pi$ contributions.%
		}
		\label{fig:ad_o_ocs_all}
	\end{figure}
	
	\begin{figure}[htb]
		\centering
		\includegraphics[width=0.95\textwidth]{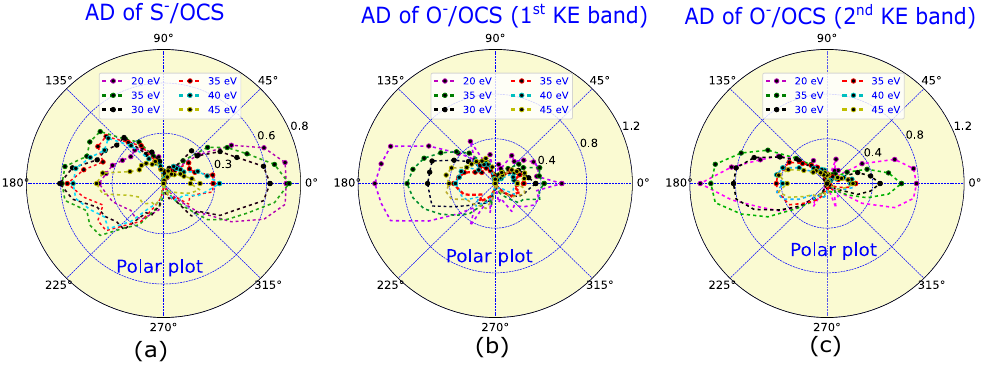}
		\caption{%
			Polar representations of angular distributions for (a)~$\Sm$ and (b,c)~$\Om$ fragments at selected incident energies.%
		}
		\label{fig:ad_ocs_polar}
	\end{figure}
	
	Figures~\ref{fig:ad_s_ocs_all} and~\ref{fig:ad_o_ocs_all} display KE-integrated angular distributions with fits to equation~(\ref{eq:vanbrunt}) using partial waves up to $l_{\mathrm{max}} = 3$. Polar plots appear in figure~\ref{fig:ad_ocs_polar}.
	
	Tables~\ref{tab:ocs_s_polar} and~\ref{tab:ocs_o_polar} collect optimised fitting parameters. At every beam energy and for both channels, $\beta$ exceeds unity ($\beta = 1.2$--$2.2$), proving that the dipole-Born approximation cannot account for electron-impact IPD of $\OCS$. This extends patterns documented for $\Otwo$, $\CO$, and $\text{CO}_2$ \cite{kundu2023breakdown,kundu2022_ipd_co,kundu2020ipd_co2}.
	
	P-wave ($l=1$) strength is most prominent between 20 and 30~eV, while F-wave ($l=3$) contributions grow above 35~eV. This shift reflects how different angular-momentum transfer channels open as the electron probes deeper regions of the molecular potential \cite{greene2000creation,swave_scatter_prl,topical_rev_green,bardsley1968resonant,eiles2019trilobites,shape_feshbach_jctc_2022,shape_reso_ulrs}.
	
	Allowing $|\mu|=1$ contributions ($\Sigma \leftrightarrow \Pi$ transitions) improves $\chi^{2}$, indicating a role for bending motion \cite{worth2004beyond}. The relative phase shifts $\delta_l$ vary smoothly with beam energy, as expected for coherent excitation.
	
	\begin{table}[htb]
		\centering
		\caption{Fitting parameters for $\Sm$ angular distributions assuming $\Sigma \rightarrow \Sigma$ transitions.}
		\label{tab:ocs_s_polar}
		\begin{tabular}{ccccc}
			\toprule
			$E_{\mathrm{beam}}$ (eV) & $a_0:a_1:a_2:a_3$ & $(\delta_{s\text{-}p},\delta_{s\text{-}d},\delta_{s\text{-}f})$ (rad) & $\beta$ & $R^{2}$ \\
			\midrule
			20 & 0.98:1.41:1.16:1.35 & (1.70, 0.76, $-$1.49) & 1.49 & 0.99 \\
			25 & 4.09:1.51:1.11:0.62 & (1.48, 1.68, 3.14)    & 1.67 & 0.98 \\
			30 & 0.87:1.38:1.15:0.70 & (1.68, 1.56, 3.14)    & 1.94 & 0.96 \\
			35 & 0.99:1.34:0.13:2.24 & (1.35, 1.54, 1.74)    & 1.68 & 0.96 \\
			40 & 0.87:1.31:0.33:2.73 & (1.21, 1.43, 1.47)    & 1.49 & 0.98 \\
			45 & 0.82:0.86:0.33:2.46 & (0.26, 0.70, $-$3.14) & 1.76 & 0.96 \\
			\bottomrule
		\end{tabular}
	\end{table}
	
	\begin{table}[htb]
		\centering
		\caption{Fitting parameters for $\Om$ angular distributions assuming $\Sigma \rightarrow \Sigma$ transitions, analysed separately for low-KE (1st) and high-KE (2nd) bands.}
		\label{tab:ocs_o_polar}
		\begin{tabular}{cccccc}
			\toprule
			Band & $E_{\mathrm{beam}}$ (eV) & $a_0:a_1:a_2:a_3$ & $(\delta_{s\text{-}p},\delta_{s\text{-}d},\delta_{s\text{-}f})$ (rad) & $\beta$ & $R^{2}$ \\
			\midrule
			1st & 20 & 0.87:1.58:1.32:0.90 & (1.12, 1.27, 3.14)     & 2.18 & 0.93 \\
			1st & 25 & 0.91:1.53:0.77:0.34 & (1.10, 1.18, 3.14)     & 1.83 & 0.99 \\
			1st & 30 & 0.86:1.39:0.46:2.99 & (0.36, 1.02, $-$2.82)  & 1.36 & 0.98 \\
			1st & 35 & 0.85:0.95:1.12:1.73 & (2.36, 1.87, $-$1.58)  & 1.50 & 0.96 \\
			1st & 40 & 0.94:1.10:0.36:1.64 & (1.78, $-$0.78, 2.11)  & 1.63 & 0.94 \\
			1st & 45 & 0.89:1.31:0.85:2.21 & (0.46, 0.85, 3.14)     & 1.51 & 0.94 \\
			\midrule
			2nd & 20 & 0.99:1.31:0.28:4.56 & (0.70, 0.04, 0.48)     & 1.55 & 0.96 \\
			2nd & 25 & 0.98:1.26:0.47:3.11 & (1.18, 1.74, 0.88)     & 1.65 & 0.99 \\
			2nd & 30 & 0.99:0.96:1.34:4.78 & (1.98, 2.44, 1.32)     & 1.21 & 0.99 \\
			2nd & 35 & 1.00:1.01:0.86:4.36 & (1.40, 1.77, 1.49)     & 1.22 & 0.98 \\
			2nd & 40 & 0.93:0.88:1.69:0.90 & (1.69, 1.46, $-$1.21)  & 1.23 & 0.96 \\
			2nd & 45 & 0.91:0.69:1.98:0.17 & (1.52, 1.53, 1.18)     & 1.20 & 0.96 \\
			\bottomrule
		\end{tabular}
	\end{table}
	
	\subsection{Angular anisotropy}
	
	We define the anisotropy parameter $\alpha = (I_{\mathrm{max}} - I_{\mathrm{min}})/(I_{\mathrm{max}} + I_{\mathrm{min}})$ \cite{zare1967dissociation}.
	
	\begin{figure}[htb]
		\centering
		\includegraphics[width=0.65\textwidth]{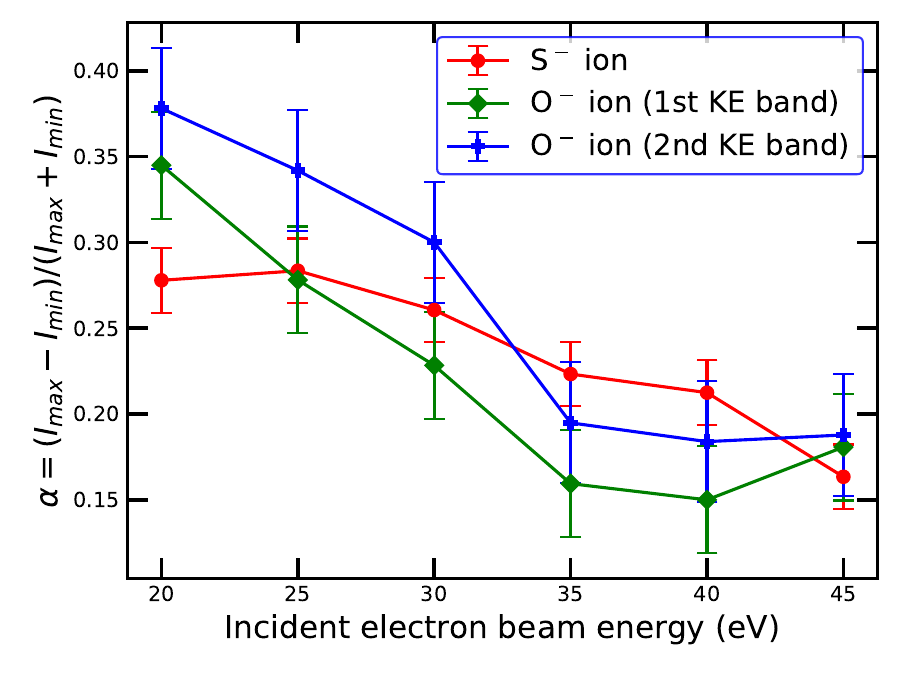}
		\caption{%
			Angular anisotropy parameter $\alpha$ as a function of incident electron energy for $\Om$ (first and second KE bands) and $\Sm$ fragments.%
		}
		\label{fig:ocs_ang_aniso}
	\end{figure}
	
	Figure~\ref{fig:ocs_ang_aniso} plots $\alpha$ against beam energy. For both channels $\alpha$ drops monotonically from roughly 0.15 at 20~eV to approximately 0.05 at 45~eV.
	
	A forward-backward asymmetry appears in the angular data (figures~\ref{fig:ad_s_ocs_all}--\ref{fig:ad_ocs_polar}): intensity at $\theta = 0^{\circ}$ differs from that at $\theta = 180^{\circ}$. This asymmetry originates in quantum interference between partial waves of opposite parity \cite{dill1976angular,reid2003photoelectron}. Cross terms between even-$l$ and odd-$l$ amplitudes inject a $\cos\theta$ modulation. That this asymmetry persists indicates coherent excitation with well-defined relative phases.
	
	\subsection{Mechanistic interpretation}
	
	The observables converge on a unified mechanistic picture for electron-impact quasi-resonant intramolecular IPD in $\OCS$. Table~\ref{tab:mechanism_summary} summarises the findings.
	
	\begin{table}[htb]
		\centering
		\caption{Summary of experimental observables and their mechanistic implications.}
		\label{tab:mechanism_summary}
		\begin{tabular}{p{3.2cm}p{4.2cm}p{5.0cm}}
			\toprule
			Observable & Observation & Mechanistic implication \\
			\midrule
			Threshold energy & Channels I and II agree with thermochemistry & Genuine intramolecular fragmentation via IPD \\
			\addlinespace
			KE vs.\ incident energy & Maximum KE saturates above 30~eV & Quasi-resonant pathway via discrete superexcited doorway states \\
			\addlinespace
			KE distribution shape & Unimodal ($\Sm$); bimodal ($\Om$) & Single $\COp$ electronic state; two $\CSp$ states ($X$ and $A$) \\
			\addlinespace
			Partial waves & P-wave dominant at low energy; F-wave at high energy & Energy-dependent angular momentum transfer \\
			\addlinespace
			$\beta$ parameter & $>1$ at all energies & Dipole-Born approximation breaks down \\
			\addlinespace
			Angular anisotropy & Small, decreasing; forward-backward asymmetric & Coherent excitation; predissociation \\
			\addlinespace
			VMI image structure & Diffuse, structureless & Predissociation through strongly coupled avoided crossings \\
			\bottomrule
		\end{tabular}
	\end{table}
	
	The data point to a \textit{quasi-resonant} route in which the impinging electron kicks $\OCS$ into discrete hybrid superexcited configurations blending Rydberg and ion-pair character (equation~\ref{eq:hybrid_state}) \cite{merkt1997molecules,merkt1998high,bardsley1968resonant}. Once the scattered electron departs, the superexcited molecule fragments on its own---the subsequent ion-pair production is therefore a \textit{unimolecular} process whose outcome is dictated solely by the internal energy and symmetry of the doorway state \cite{baer1996unimolecular,steinfeld1999chemical}. The saturation of maximum KER above 30~eV provides the clearest experimental fingerprint: the product energy budget is capped by the energetics of the discrete intermediate, irrespective of the incident electron's kinetic energy.
	
	Unlike statistical (RRKM-type) unimolecular decay \cite{marcus1952unimolecular,marcus1965rrkm_theory}, the structured KER distributions and coherent partial-wave signatures show that complete intramolecular vibrational redistribution does not precede fragmentation \cite{nesbitt1996ivr_review,gruebele2004ivr}. The observations are consistent with \textit{state-specific unimolecular dissociation} mediated by nonadiabatic transitions at avoided crossings \cite{crim1996statespecific,schinke1993photodissociation}. Direct state identification would require coincidence measurements or high-resolution electron energy loss spectroscopy.
	
	In the Fermi pseudopotential language \cite{fermi1934sopra,omont1977theory}, these hybrid superexcited states are arrangements in which a loosely bound Rydberg electron interacts at short range with the ionic core. As the molecule slides along the dissociation coordinate, the ion-pair weight of the wavefunction grows at each successive avoided crossing until intramolecular charge separation is complete \cite{kirrander2018heavy,greene2000creation}.
	
	The two KE humps of $\Om$ indicate that Channel~II taps doorway states correlating with both $\CSp$($X\,^{2}\Sigma^{+}$) $+$ $\Om$ and $\CSp$($A\,^{2}\Pi$) $+$ $\Om$ limits \cite{suzuki1998nonadiabatic,mcglynn1971electronic}. The single hump of $\Sm$ indicates Channel~I runs primarily through states connecting to $\COp$($X\,^{2}\Sigma^{+}$) \cite{huber1979constants}.
	
	Electron impact can also reach optically dark configurations---states lacking a dipole transition moment from the ground state but accessible through exchange scattering \cite{allan2007electron,khakoo2005dark}. For $\OCS$ the relevant dark states include $a\,^{3}\Pi$ and $A'\,^{3}\Sigma^{+}$ triplets \cite{suzuki1998nonadiabatic}. Spin-orbit-driven intersystem crossing from these triplets into ion-pair continua could contribute to the observed IPD \cite{marian2012spin,rothbaum2026ultrafast}.
	
	\subsection{Relationship to interatomic Coulombic electron capture}
	
	Interatomic Coulombic electron capture (ICEC) \cite{gokhberg2010icec_theory,jahnke2020icec_icd_review,trinter2014icec_exp,averbukh2004_icec} provides an alternative conceptual framework for ion-pair formation: an incoming electron is captured at one atomic site while the released energy ejects an electron from a neighbouring site \cite{cederbaum1997giant}. Although ICEC has been studied primarily in weakly bound complexes where perturbative intersite energy transfer applies \cite{wehrli2021charge_pre}, the underlying physics---electron capture mediated by Coulomb coupling---shares features with the hybrid Rydberg--ion-pair states invoked here.
	
	Several observations \cite{chakraborty2016dipolar,larsen_dark,kundu2022_ipd_co,mitsuke1993negative} suggest that the present measurements are more naturally described by intramolecular nonadiabatic IPD than by ICEC. First, our room-temperature effusive beam conditions disfavour dimer or cluster formation. Second, the measured thresholds agree with bond-dissociation thermochemistry. Third, the cross sections rise monotonically from threshold and plateau above 30~eV---behaviour characteristic of quasi-resonant excitation through discrete doorways \cite{bardsley1968resonant}---rather than displaying the resonance structure expected for ICEC. These kinematic signatures---thresholds matching thermochemistry, clean two-body Newton spheres in VMI, and smooth excitation functions without sharp resonances---all confirm genuine intramolecular fragmentation rather than intermolecular cluster breakup or ICEC-mediated three-body dissociation.
	
	That said, as the molecule stretches along the dissociation coordinate, the distinction between intramolecular charge redistribution and ICEC-like intersite transfer becomes less sharp. Coincidence measurements correlating the scattered electron energy and angle with fragment momenta would be needed to quantify any ICEC contribution and to map the transition from strong-coupling (intramolecular) to weak-coupling (ICEC-like) regimes. For the present data, the intramolecular nonadiabatic IPD framework provides a consistent and sufficient explanation.
	
	\section{Conclusions}
	
	We have carried out a velocity map imaging study of electron-impact quasi-resonant intramolecular ion-pair dissociation of carbonyl sulfide, resolving two channels: $\COp + \Sm$ (Channel~I, threshold $14.8 \pm 0.7$~eV) and $\CSp + \Om$ (Channel~II, threshold $16.8 \pm 0.7$~eV). Both thresholds agree with thermochemical predictions and earlier photoionisation data.
	
	The velocity slice images are diffuse and featureless, as expected when predissociation runs through tightly coupled avoided crossings. Kinetic energy release spectra are single-peaked for $\Sm$ (maximum KE $\sim$2.0~eV) and double-peaked for $\Om$ (maximum KE $\sim$3.0~eV). Within the heavy Rydberg framework, the bimodal structure of Channel~II arises from doorway states correlating with different electronic levels of $\CSp$ ($X\,^{2}\Sigma^{+}$ and $A\,^{2}\Pi$).
	
	The levelling-off of maximum kinetic energy above 30~eV constitutes the strongest evidence for quasi-resonant IPD: fragmentation proceeds through discrete superexcited doorway states rather than direct continuum excitation \cite{merkt1997molecules,hatano2001interaction,bardsley1968resonant}. Combined with the structured KER distributions and coherent partial-wave signatures, this saturation demonstrates that the process is consistent with state-specific unimolecular dissociation rather than statistical decay.
	
	Partial-wave fitting confirms that the dipole-Born approximation fails ($\beta > 1$) at all energies and uncovers a systematic shift in the partial-wave mix---P-wave prominence at 20--30~eV giving way to appreciable F-wave weight at 35--45~eV.
	
	The data support a quasi-resonant mechanism in which electron impact populates hybrid superexcited states of mixed Rydberg--ion-pair character, which then fragment nonadiabatically through intramolecular charge redistribution at strongly coupled avoided crossings. Alternative pathways such as interatomic Coulombic electron capture are inapplicable given the compact covalent geometry of isolated $\OCS$ molecules.
	
	These results sharpen our understanding of how electrons drive molecular fragmentation, a question of practical importance given that $\OCS$ is the leading sulfur-bearing trace species in the stratosphere and a key player in stratospheric aerosol formation \cite{ueno_young_sun_para,hattori2020constraining}. The energy-dependent branching ratios and partial-wave compositions reported here provide benchmark data for modelling electron-initiated chemistry in planetary ionospheres and plasma environments.
	
	\section*{Acknowledgments}
	
	We acknowledge financial support from the Department of Science and Technology, Government of India, and the Indian Institute of Science Education and Research Kolkata. NK thanks the INSPIRE Fellowship for financial support.
	
	\section*{Author contributions}
	
	NK and DN conceived the experiment. NK and SG performed the measurements and data analysis. NK wrote the manuscript with input from all authors. DN supervised the project.
	
	\section*{Conflicts of interest}
	
	The authors declare no conflicts of interest.
	
	\section*{Data availability statement}
	
	The data supporting the findings of this study are available from the corresponding author upon reasonable request.
	
	\bibliography{cite}
	
\end{document}